# Dodecagonal quasicrystal in Mn-based quaternary alloys containing Cr, Ni and Si

Shuhei Iwami and Tsutomu Ishimasa*

*Graduate School of Engineering, Hokkaido University, Kita-ku, 060-8628 Sapporo, Japan*

**Abstract**

A dodecagonal quasicrystal showing 12-fold symmetry forms in Mn-rich quaternary alloys containing 5.5 or 7.5 at.% Cr, 5.0 at.% Ni and 17.5 at.% Si. After annealing at 700 ºC for 130 h, the quasicrystal precipitated in a matrix of β-Mn type crystalline phase. The shape of the quasicrystal is needle-like having a length of several tens of micrometers. Electron diffraction as well as powder X-ray diffraction experiments have revealed the following characteristics of the quasicrystal; diffraction symmetry 12/*mmm*, presence of systematic extinction for $h_1h_2h_2h_1h_5$ type reflections with odd $h_5$ index, and then five-dimensional space group *P*12$_6$/*mmc*. Indexing of the reflections indicated that the dimension of the common edge in the equilateral triangle-square tiling is 4.560 Å, and the periodicity is 4.626 Å along the 12-fold axis. This is the first example of the dodecagonal quasicrystal synthesized by ordinary metallurgical method in 3-*d* transition-metal alloys.

*Corresponding Author

E-mail : ishimasa@eng.hokudai.ac.jp

Telephone:   +81-11-706-6643

Fax:         +81-11-716-6175

1. **Introduction**

A dodecagonal quasicrystal showing 12-fold diffraction symmetry is very rare in the area of alloys, contrary to the other types of quasicrystals such as icosahedral and decagonal. In particular, in the case of 3-*d* transition-metal alloys, only three examples have been reported; namely a Ni-Cr small particle synthesized by gas-evaporation technique [1], rapidly cooled Ni-based alloys [2], and a Bi-Mn film made by vacuum deposition [3]. There is no report on the formation of a dodecagonal quasicrystal using ordinary metallurgical method; *i.e.* melting, solidification and annealing. This may be due to lack of phase stability of a dodecagonal quasicrystal, although some theoretical works suggested its stability [4, 5, 6].

In the course of the research on icosahedral quasicrystals, it has been recognized that an approach starting from an approximant crystal is very effective to find a new quasicrystal alloy. Here an approximant crystal is a periodic structure with a local atomic arrangement called cluster similar to that present in the corresponding quasicrystal. In this letter, it will be shown that such an approach is also applicable to the research of a dodecagonal quasicrystal.

A dodecagonal quasicrystal has a unique 12-fold symmetry axis and periodicity along that axis. In the plane perpendicular to the 12-fold axis, the ideal dodecagonal quasicrystal may be expressed as a quasiperiodic tiling of equilateral triangle and square with the common edge length *a*. Such a model, sometimes including 30º rhombus, has been discussed by several authors [7, 8, 9, 10, 11]. In order to construct three-dimensional model of the real dodecagonal quasicrystals, an atomic decoration model of the two-dimensional tiling has been proposed, where a unit cell of the cubic A15-type ($Cr_3Si$-type) structure is assigned to the square, and a half unit cell of the hexagonal $Zr_4Al_3$-type to the triangle [1, 12]. In this model, the atoms are situated in four layers located at $z = 0$, $\pm 1/4$ and $1/2$ along the 12-fold axis. Each vertex of the triangle-square tiling is decorated by a hexagonal anti-prism of which projection along the 12-fold axis is a dodecagon. Gähler studied this type of structure model as a projection of a five-dimensional periodic structure, and predicted characteristic extinction in a diffraction pattern [10].

This decoration model is based on the structure of the σ-phase that is a tetragonal crystal appearing in many transition metal alloys. In the *c*-projection of the σ-phase structure, one can find triangle-square tiling of the tesselation $3^2 \cdot 4 \cdot 3 \cdot 4$. Then the σ-phase is regarded as an approximant crystal to the dodecagonal quasicrystal. Besides this, a series of approximant crystals is known in Ni-based superalloy [13]. Additionally a large hexagonal approximant crystal of the dodecagonal quasicrystal has been discovered in Mn-Cr-Si and Mn-V-Si systems, of which composition is $Mn_{70}(Cr, V)_{10}Si_{20}$ [14, 15, 16]. Its unit cell consists of seven triangles and three squares, which is regarded as the second generation triangle scaled by the factor of $2 + \sqrt{3}$.

In this letter, we will report experimental results on a new dodecagonal quasicrystal formed in quaternary Mn-based alloys containing Cr, Ni and Si, which may be related to the hexagonal approximant $Mn_{70}Cr_{10}Si_{20}$.

## 2. Experimental procedures

Specimens of Mn-based alloys were prepared using high-purity materials of Mn (Nilaco; 99.9+ %), Cr (Alfa Aesar; 99.996 %), Ni (Nilaco; 99.99 %) and Si (Rare metallic; 99.999+ %). In this letter, we will describe the experimental results on the following two alloys with slightly different compositions: $Mn_{72.0-x}Cr_{5.5+x}Ni_{5.0}Si_{17.5}$ with the parameter $x = 0.0$ and 2.0. They were synthesized using an electric arc furnace in an atmosphere of Ar. In this process an alloy ingot was melted three times on the water-cooled copper hearth. They are brittle and tend to break into few pieces. The weight loss in this melting process was less than 1 % except for the loss due to the breaking. A specimen was sealed in a silica ampoule after evacuating to a pressure of approximately $2.0 \times 10^{-6}$ Torr. The specimens were annealed at 700 °C for 130 or 230 h. After the annealing, they were cooled in water without breaking the silica ampoules. The weight loss during the heat treatment was less than 0.1%. Hereafter the specimens will be denoted by their nominal compositions.

The phases included in each specimen were analyzed by means of X-ray diffraction method using Cu-Kα radiation with a Rigaku RINT-2000 diffractometer. According to [11] in the case of dodecagonal quasicrystal, a reciprocal vector $g_{//}$ is expressed by the following formula;

$$g_{//} = \sum_{n=1}^{5} h_n a_n^* \quad\quad --- (1),$$

$$a_n^* = \frac{1}{\sqrt{3}a}\left\{i\cos\frac{(n-1)\pi}{6} + j\sin\frac{(n-1)\pi}{6}\right\} \quad \text{for } n = 1\sim 4,$$

$$a_5^* = \frac{1}{c}k .$$

Here three vectors $i$, $j$ and $k$ denote three-dimensional unit vectors orthogonal to each other, and the vectors $a_1^* \sim a_4^*$ span on the quasiperiodic plane with the indices $h_1 \sim h_4$. The vector $a_5^*$ is parallel to the 12-fold axis. Parameters $a$ and $c$ in the direct space denote the common edge length of the equilateral triangle and the square, and the period along the 12-fold axis, respectively. These parameters were determined by modified Cohen's method for the peak positions listed in Table I. Because the dodecagonal quasicrystal always coexists with the dominant β-Mn type phase [17] in this study, subtraction of the contribution of the latter phase from a measured X-ray diffraction pattern was carried out after precise profile decomposition by assuming pseudo-Voigt function for each reflection. A lattice parameter of a cubic β-Mn type phase was determined by Cohen's method.

Selected-area electron diffraction patterns were observed by a JEOL JEM200CS microscope equipped with a double tilting stage at the acceleration voltage 200 kV. Composition analysis was carried out by a JEOL JXA-8530 electron probe microanalyzer after detection of domains of each phase by observing back-scattered electron images. An average value of composition of each phase was estimated from more than ten measurements at different positions in the samples. It was estimated that the spatial fluctuation of the composition is less than 1.0 at.% inside each phase.

## 3.   Results and Discussion

Figure 1 shows typical selected-area electron diffraction patterns of the dodecagonal quasicrystal formed in the $Mn_{72.0-x}Cr_{5.5+x}Ni_{5.0}Si_{17.5}$ alloy with $x = 0.0$ annealed at 700 ºC for 130 h. It was determined that the diffraction symmetry is 12/*mmm* because of the presence of a 12-fold axis in Fig. 1 (a), and two kinds of 2-fold axes both perpendicular to the 12-fold axis. The diffraction patterns in Figs. 1(b) and (c) correspond to these 2-fold axes, and are related to each other by 15 degrees rotation around the 12-fold axis. While some diffuse streaks are present, all reflections observed in these patterns can be indexed using the equation (1). This is evidence of the formation of a dodecagonal quasicrystal in this alloy. From these diffraction patterns, the measured *a*/*c* ratio was 0.982(2). This value was further confirmed by means of powder X-ray diffraction.

The scaling property by the factor of $2 + \sqrt{3}$ characteristic to the dodecagonal quasicrystal is noticed, and examples are indicated by arrows at lower parts of Figs. 1(b) and (c). In Fig. 1(b) short and long arrows correspond to 0 1 0 0 0 and 1 2 1 0 0 reflections, respectively. In Fig. 1 (c), three arrows correspond to the following three reflections: -1 1 1 -1 0, 0 1 1 0 0 and 1 3 3 1 0. The former two reciprocal lengths as well as the latter three are related by the scaling with the factor of $2 + \sqrt{3}$.

By comparing two diffraction patterns in Figs. 1(b) and (c), it is noticed that the reflections on the line with odd $h_5$ are not visible in (c) except for faintly visible 0 0 0 0 1 reflection. This indicates systematic extinction with respect to $h_1h_2h_2h_1h_5$ type reflections with odd $h_5$. As an example, the arrowhead in Fig. 1(c) indicates position of 1 2 2 1 1 reflection which is not visible. This systematic extinction is due to the presence of a $12_6$-screw axis as well as a glide plane [10]. Referring to Table 10 and its caption in [11], the five-dimensional space group of this dodecagonal quasicrystal is assigned to $P12_6/mmc$. In Fig. 1(c) one may notice weak diffuse streak along the line $h_1h_2h_2h_1h_5$ with odd $h_5$, and also very weak spot at 0 0 0 0 1. The appearance of 0 0 0 0 1 reflection seems to violate the extinction rule, but on the line it does not show all other reflections. Therefore this occurrence is considered not to be essential, and is possibly caused by multiple diffraction effect through higher Laue zone and/or by slightly misaligned incident beam.

Electron microscopy observation has also revealed existence of small amount of a hexagonal crystal in the present $x = 0.0$ alloy. This hexagonal crystal is the approximant equivalent to that observed in Mn-Cr-Si system [15, 16]. However, in the X-ray diffraction pattern, any reflection caused by this crystal was not detected because of its small quantity.

Figure 2 summarizes the result of powder X-ray diffraction experiment. The X-ray diffraction pattern of the annealed alloy with $x = 0.0$ consists of two components; one is β-Mn type phase with the lattice parameter $a_\beta = 6.2688$ (3) Å, and the other the dodecagonal quasicrystal. The detailed structure of the β-Mn type was already reported in [17]. Therefore it was easy to subtract the contribution of this phase, shown in Fig. 2(b), from Fig. 2(a). The result is presented in Fig. 2(c). Referring to the indexing scheme (1) and the observed electron diffraction patterns, the reflections in Fig. 2(c) were indexed as a dodecagonal quasicrystal, while small artifacts appeared during the subtraction treatment. The reflections of the quasicrystals are summarized in Table I, which again satisfy the extinction rule. It was determined that the parameters $a$ and $c$ characteristic to the dodecagonal quasicrystal are 4.560 (6) Å and 4.626 (8) Å, respectively. The value $a/c = 0.986$ agrees with the one measured by the electron diffraction method.

These values are consistent to the dimensions of triangle- and square-bricks embedded in the hexagonal Mn-Cr-Si approximant with $a_h = 16.985$ Å and $c_h = 4.625$ Å [15]. Here it is calculated that the corresponding parameters are $a = a_h/(2 + \sqrt{3}) = 4.551$ Å and $c = c_h = 4.625$ Å. In the structure model of the hexagonal phase, the triangle- and square-bricks have inner atomic arrangement corresponding to the $Zr_4Al_3$-type and A15-type structures, respectively. A structure model of the dodecagonal quasicrystal with this type of atomic decoration was studied, and its electron diffraction patterns were calculated in [10]. They are presented in Fig. 3, which are compared with the experimental ones shown in Fig. 1. In this calculation, multiple diffraction effect was treated using Darwin's method. Over all agreement between the calculated and observed ones is remarkable. Especially the systematic extinction observed in Fig. 1(c) is reproduced in the calculated one in Fig. 3(c). This agreement supports the validity of the atomic decoration with $Zr_4Al_3$ and A15-types. This structure model is now examined by means of high-resolution electron microscopy, and the result will be described elsewhere.

Back-scattered scanning electron micrograph of the annealed alloy with $x = 0.0$ is presented in Fig. 4. Needle-like precipitates having a length of several tens of micrometers are observed in a grey matrix. Alloy compositions of the precipitate and the matrix were determined to be $Mn_{73}Cr_3Ni_7Si_{17}$ and $Mn_{74}Cr_{10}Ni_1Si_{15}$, respectively. In the literature [17], it was reported that the composition of β-Mn type phase is $Mn_{63}Ni_{20}Si_{17}$, and so we assign the former composition with higher Ni and Si contents to the β-Mn type structure, while the needle-shaped precipitates with $Mn_{74}Cr_{10}Ni_1Si_{15}$ composition containing only 1 at.% Ni to the quasicrystal. This assignment is also supported by the fact that the X-ray diffraction pattern in Fig. 2(a) indicates small quantity of the quasicrystal in this alloy.

Powder X-ray diffraction experiment revealed that at the same composition with $x = 0.0$, the as-cast alloy without additional heat treatment includes the β-Mn type exclusively, while the dodecagonal quasicrystal forms during the annealing process at 700 ºC. This fact suggests the phase stability of the dodecagonal quasicrystal at 700 ºC at this composition. The identification of the stability, namely thermodynamically stable or metastable, is the important problem that we must clarify in the future.

At the different composition with $x = 2.0$, the as-cast alloy included small amount of σ-phase in addition to the dominant β-Mn type phase. Annealing at 700 ºC for 230 h caused the formation of the dodecagonal quasicrystal as in the case of the former alloy with $x = 0.0$. Both σ- and β-Mn type phases remained after the annealing, coexisting with the dodecagonal quasicrystal. Probably due to this complicated situation, highly defected σ-phase was observed, and the dodecagonal quasicrystal was less uniform. Namely degree of structural perfection of the dodecagonal quasicrystal depends on the place in the specimen as judged from the quality of the electron diffraction patterns.

## 4. Conclusion

A new dodecagonal quasicrystal belonging to the five-dimensional space group $P12_6/mmc$ has been observed in the quaternary Mn-based alloys containing Cr, Ni and Si. This quasicrystal forms by heat treatment at 700 ºC, and then it has better stability than other dodecagonal quasicrystals previously reported in 3-$d$ transition-metal alloys. Owing to this fact, it is now possible to observe complete set of electron diffraction patterns as well as powder X-ray diffraction pattern providing insight into detailed structure of this novel state of matter.

**Acknowledgement**

The authors thank Marek Mihalkovič for valuable discussions and advice in the manuscript preparation, and Hayato Iga and Yuya Tanaka for their contributions to this research project. They also thank Nobuyuki Miyazaki for his help in the use of JXA-8530 electron probe microanalyzer.

**Figure Captions**

Figure 1

Selected-area electron diffraction patterns exhibiting diffraction symmetry 12/*mmm*. (a) along 12-fold axis, (b) along 2-fold axis, (c) along another 2-fold axis. The indices of reflections are A: 1 2 1 0 0, B: 2 3 2 0 0, C: 1 2 2 1 0, D: 1 3 3 1 0 and E: 0 0 0 0 2. Arrowheads indicates position of 1 2 2 1 1 reflection which is not visible.   Arrows in lower parts of (b) and (c) illustrate $2 + \sqrt{3}$ scaling.

Figure 2

(a) Measured powder X-ray diffraction pattern of $Mn_{72.0}Cr_{5.5}Ni_{5.0}Si_{17.5}$ alloy annealed at 700 ºC for 130 h. Arrows indicate reflections due to dodecagonal quasicrystal, while two of them at 78.42º and 81.18º can be seen only in (c). (b) Intensity distribution of β-Mn type phase. All reflections are indexed as a simple cubic structure with $a = 6.2688$ Å. (c) After subtraction of the contribution of β-Mn type phase from (a). In the vertical axis, it is extended 5 times to be easy to look. Bars indicate reflections of the dodecagonal quasicrystal.

Figure 3

Simulated electron diffraction patterns of dodecagonal quasicrystal reproduced with permission [10]. (a) Dodecagonal plane. (b) Mirror plane. (c) Glide plane. Notice systematic extinction in (c) that corresponds to observed one in Fig. 1(c).

Figure 4

Back-scattered scanning electron micrograph (composition image) of $Mn_{72.0}Cr_{5.5}Ni_{5.0}Si_{17.5}$ alloy annealed at 700 ºC for 130 h. Dark needle and gray matrix correspond to dodecagonal quasicrystal and β-Mn type, respectively.

Table 1. Reflections and indices. For the calculation of angle $2\theta$, the parameters $a = 4.560$ Å and $c = 4.626$ Å were used.

| Measured $2\theta$ (degree) | Calculated $2\theta$ (degree) | Index | | | | |
|---|---|---|---|---|---|---|
| 38.90 | 38.90 | 0 | 0 | 0 | 0 | 2 |
| 42.64 | 42.69 | 2 | 2 | 0 | -1 | 0 |
| 44.98 | 44.99 | 1 | 1 | 0 | 0 | 2 |
| 47.12 | 47.19 | 2 | 2 | 0 | -1 | 1 |
| 50.36 | 50.49 | 1 | 2 | 0 | -1 | 2 |
| 75.52 | 75.62 | 2 | 3 | 1 | -1 | 2 |
| 76.38 | 76.35 | 2 | 2 | 0 | -1 | 3 |
| 78.42 | 78.16 | 3 | 4 | 0 | -2 | 0 |
| 81.18 | 81.39 | 3 | 4 | 0 | -2 | 1 |
| 83.46 | 83.52 | 0 | 0 | 0 | 0 | 4 |

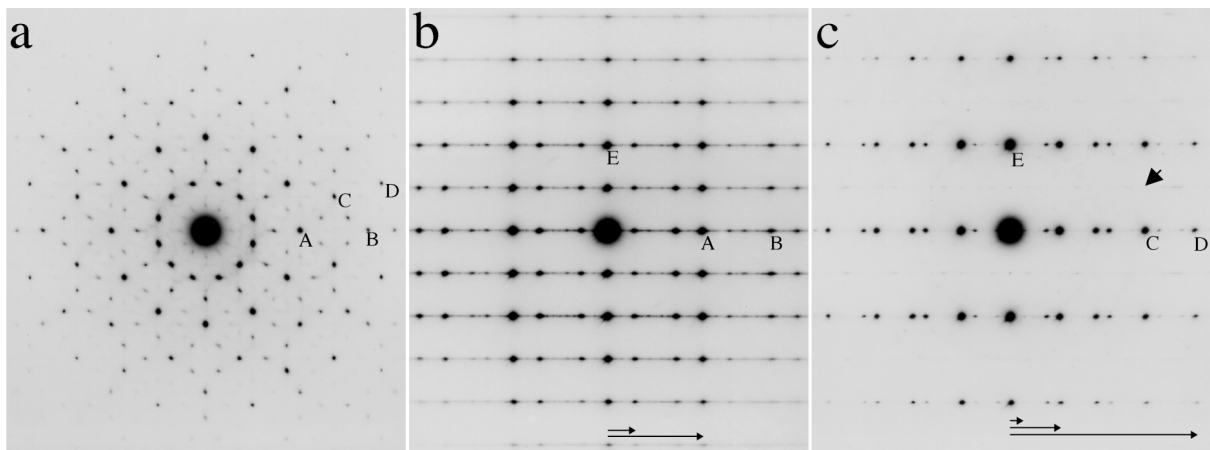

Figure 1

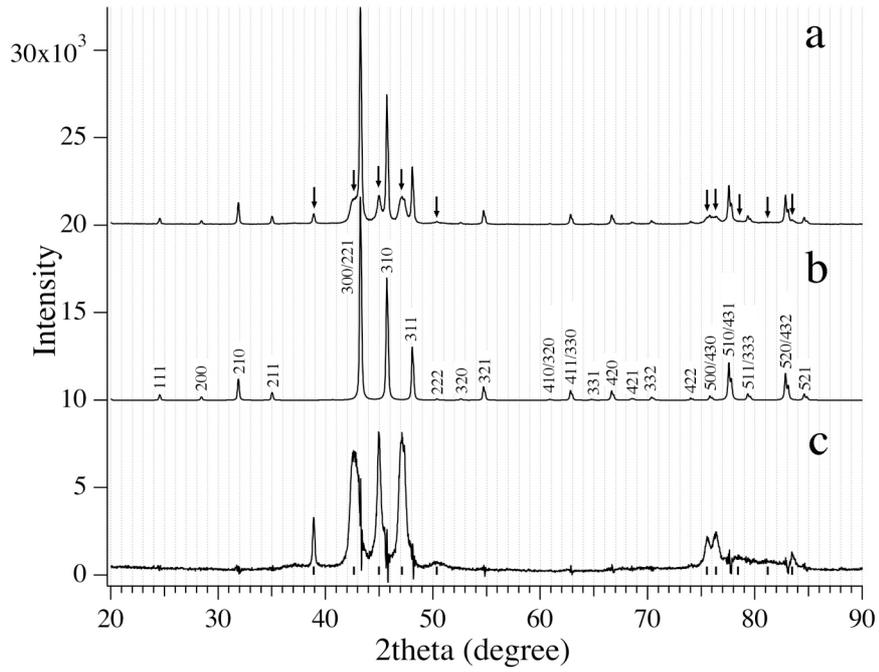

Figure 2

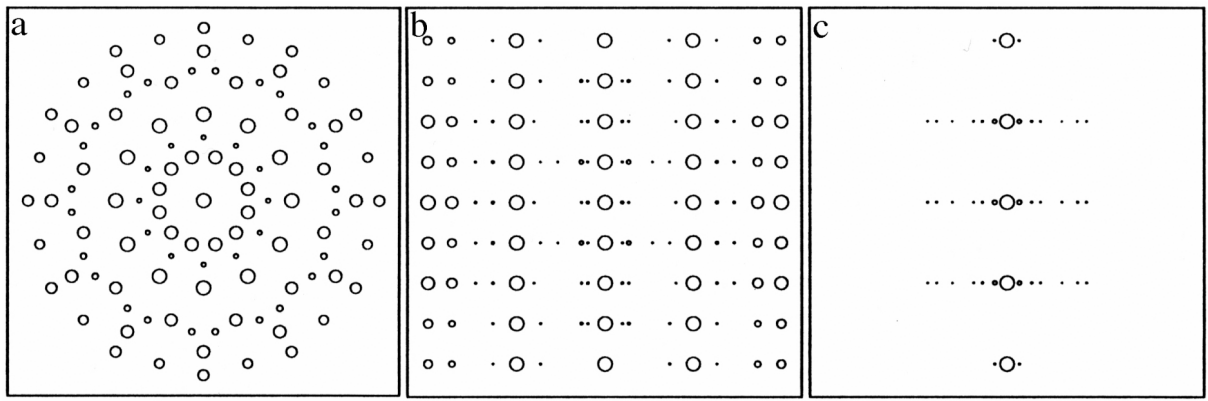

Figure 3

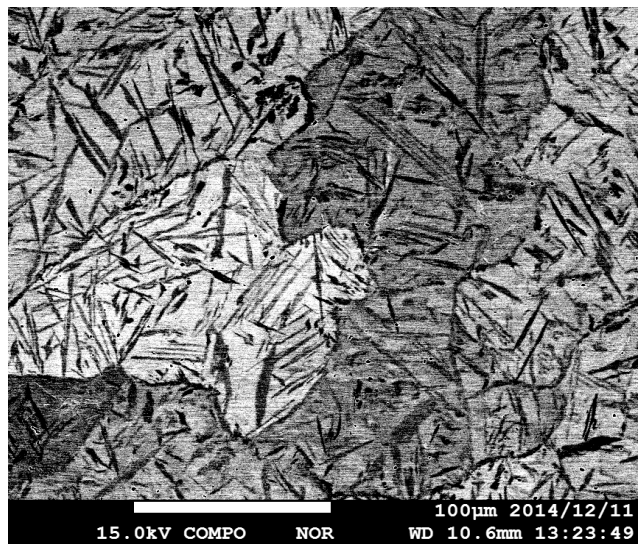

Figure 4